\documentstyle[prb,aps,psfig]{revtex}
\draft
\begin{document}
\title{Scaling of transition temerature and $CuO_2$
plane buckling in the cuprate superconductors}
\author{Manas Sardar and Satadeep Bhattacharjee}
\address{Material Science Division, I.G.C.A.R, kalpakkam 603102}
\date{\today}
\maketitle
\begin{abstract}

 A universal feature of all high temperature superconductors is the
existence of a chemical composition that gives maximum $\rm{T}_c$,
separating so called under and over-doped region. 
Recently it is seen\cite{Nature}
 that both $\rm{T}_c$ and the buckling angle of the $CuO_2$
planes goes through a maximum at the same doping level. We show that
only for optimal doping concentration the Fermi surface touches the
M($0,\pi$) point in the BZ, where the matrix element for interlayer
pair tunneling amplitude is largest, so that the gain in
delocalization energy by 
tunneling(in pairs) along the $c$ axis is largest.
Buckling of the planes on the other hand modulates the
separation between the planes and thereby modulates the interlayer
pair tunneling amplitude. That is why both $\rm{T}_c$ and buckling angle(
Oxygen atom displacement out of the plane) scales the same way with
doping concentration. We have calculated $\rm{T}_c$ and buckling angle
for various doping concentration. The agreement with experiment
is remarkably good. We also point out the possible reason for
large(about 1 percent) change
of the buckling mode phonon frequency, accross the transition
temperature for optimal doping concentration, observed in neutron
scattering
experiments\cite{Pyka}
\end{abstract}

\pacs{PACS : 63.20.Kr, 74.20.Mn,74.25.Kc,61.12 Ex, 4.72.Bk}

\section{Introduction}
 Since the discovery of High $\rm {T}_c$ cuprates a wide variety of
experiments have
shown
  strong evidence of a link between anomalies in lattice structure and
the onset of
  superconductivity in these materials. For a review of experiments
and analysis please see
  \cite{first}.
  Ion channeling experiments\cite{haga} show anomalous changes in the
oscillation amplitudes of Cu
  and O atoms at and above $\rm{T}_c$. Ultrasonic measurements\cite{noha}
shows anomalies in elastic constants.
  Infrared reflectivity measurements\cite{reed} of high quality
samples have shown
  clear connection between c-axis phonon modes and the charge
carriers.

Recently it is seen\cite{Nature} that both $\rm{T}_c$ and the buckling
angle of the $CuO_2$ planes goes through a maximum as the doping is
varied. The buckling corresponds to a static displacement of the
oxygen atoms out of the plane, with $u_{i\pm x}=-u_{i\pm y}$, where
$u_{i\pm x}$  is the displacement of the oxygen atoms at $i\pm x$
along the $c$ axis, $i$ being the $Cu$ atom position.

In inelastic neutron scsaterring experiments\cite{Pyka} it is also
seen that the buckling mode phonon  
 at 340 cm $^{-1}$ ( called $A_{1g}$ )
softens by more than 1 \%  accross the transition temperature.
 On the other hand, other modes having 
frequency close to the 340 cm$^{-1}$ mode
 like $A_u$ ( 307 cm$^{-1}$ ) and $E_2u$ ( 347 cm$^{-1}$)
 and $B_{1g}$ (285 cm $^{-1}$) are studied in detail by
 inelastic neutron scattering experiments\cite{Pyka}.
 The $A_{1g}$ mode corresponds to the vibrations of the planar
 oxygen atoms out of the plane, and out of phase with the
 adjacent CuO plane(buckling mode).
The $B_{1g}$ ( here the
oxygen atoms in two adjacent planes vibrate in phase and
out of the plane), and the  $E_{2u}$ mode( here the atoms vibrate in
the plane )
shows much less softening ( estimated to be less than .4\% and .3\%
respectively\cite{Pyka}). This compares well with the optical data
though for the
$E_{2u}$ there is no optical data available to compare with.

In this paper we shall be exploring the phenomena of
static buckling of the planes as well as the the  phonon mode
softening
within the interlayer tunneling mechanism of Wheatley, Hsu and
Anderson(WHA)\cite{wha}
and its more refined version proposed by Chakraborty et al\cite{cha}.
We will show that $T_c$ and the buckling angle varies with doping
concentration the same way. This and the  large  softening of the buckling mode
phonon  can all be understood in a unifying way.
The main ingradient of this mechanism is that the normal state is a
non
Fermi liquid and hence the single particle hopping out of planes
becomes
incoherent.
However coherent propagation of pairs of particles becomes favourable
at low enough temperatures leading to
coherent pair fluctuation and the consequent superconductivity. The
induced
Josephson coupling between the planes, or the amplitude for pair
tuneling
is quadratic in $t_{\perp}$ (bare single particle hopping amplitude
between the planes).

Any static displacement {\rm or} 
dynamic fluctuations of the atoms in the unit cell that modulates
the
distance between the planes effectively modulates the pair tunneling
amplitude between the planes. 
The main thesis of our paper,
being,
since the $A_{1g}$ mode effectively brings the planes closer together
leading to a decrease of the hopping paths for the electrons between
the
planes. This in turn makes the Josephson coupling stronger, the
electronic
system gains energy by condensation and the 
corresponding optical phonon mode($q=0$)
frequency gets softer, leading also to an average static buckling.
We find that the greater part of the large frequency shift observed
for
the $A_{1g}$ mode(buckling mode)
 is due to the screening by interlayer
pair propagation , or due pair current-phonon coupling.
The $B_{1g}$ mode on the other hand, being in phase oscillations of
the atoms keeps the distance between the planes unchanged and so it
does not couple with the pair-current flow accross the planes.

\section{Interlayer tunneling model}

We begin by writing the gap equation from interlayer tunneling
\cite{cha},
\begin{equation}
\Delta_k  =  T_J(k) {\Delta_k \over 2E_k}
{\rm tanh} {\beta E_k \over 2} + ~ \sum_{k^{\prime}}^{\prime}
V_{kk^{\prime}}~
{\Delta_k^{\prime} \over 2E_{k^{\prime}}} {\rm tanh}
{\beta E_{k^{\prime}} \over 2},
\end{equation}
This equation can be obtained by considering two close Cu-O layers as
in
BI 2212 coupled by a Josephson tunneling term of the form
$$
H_J~~ = ~~ -{1 \over t}
\sum_{k}~~t^{2}_{\perp}(k)~~(~c^{\dag}_{k\uparrow}
c^{\dag}_{-k\downarrow} d_{-k\downarrow} d_{k\uparrow} + {\rm
h.c.}~)~~,
$$
where $t_{\perp}(k)$ is the bare single electron hopping term between
the two coupled layers $c$ and $d$ and $t$ is a band structure
parameter
in the dispersion of electrons along the Cu-O plane.  Finally,
$T_J(k)$
in the right hand side of equation (1) is given by $T_J(k) =
{t^2_{\perp}(k)
\over t}$.
where $t_{\perp}(k)={t_{\perp}\over 4}({\rm cos}(kx)-{\rm
cos}(ky))^2$.
The dispersion of electrons along the Cu-O plane is chosen
to be of the form
\begin{equation}
\epsilon (k) ~~ = ~~ -2t~(~cos k_x~+cos k_y~)~~+~~
4t^{\prime}~cosk _x~cos k_y~~-2t^{\prime \prime}(\cos 2k_x+\cos 2k_y),
\end{equation}
with $t$ = 0.25 eV ,${t^{\prime} \over t}$ = 0.45 $, {t^{\prime
\prime}\over
t}=0.2$. We also choose
$\epsilon_F$ = -0.45 eV corresponding to a Fermi surface which is
closed
around the $\Gamma$ point.
These choices
are inspired by band structure calculations \cite{band}.  Note that
the
Josephson coupling term in $H_J$ conserves the individual momenta of
the
electrons that get paired by hopping across the coupled layers.  This
is
as opposed to a BCS scattering term  which would only conserve the
center of mass momenta of the pairs.  This is the origin of all
features
that are unique to the interlayer tunneling mechanism.
This term has a local $U(1)$ invariance in $k$-space and cannot by
itself give a finite ${\rm T}_c$. It needs an additional BCS type non local
interaction in the
planes which could be induced by phonons or residual correlations.
Here we
assume the inplane pairing interaction to be d-wave kind i.e,
$V_{kk^{\prime}}=V_0~f_kf_{k^{\prime}}$
with $f_k=\cos k_x -\cos k_y$. $T_J(k)$ can be infered from electronic
structurecalculations. As shown in reference\cite{band}, it is
adequate to choose $t_{\perp}(k)=
{t_{\perp}^2\over 4}(\cos k_x -\cos k_y)^2$.
According to Anderson, it is the $k$-space locality that leads to a
scale of
$\rm{T}_c$ that is linear in interlayer  pair tunneling
matrix element. He finds that in the limit $T_J > V_{kk^{\prime}}$,
$k_BT_c
\approx {T_J\over 4}$ and in the other limit, $k_B T_c\approx \hbar
\omega_D
e^{-{1\over \rho_0 V_0}}$, where $\omega_D$ $\rho_0$ and $V_0$ are
Debye
frequency, density of states at the fermi energy, and fermi surface
average matrix
element $V_{kk^{\prime}}$.
The important point being that even with a little help from the in
plane
pairing interaction the interlayer tunneling term can provide a large
scale
of $\rm{T}_c$.
In our naive analysis with a fixed cut off $\omega_D$
 we find that the $\rm{T}_c$ drops when the fermi surface
 moves away from the $M(0,\pi)$ point in a parabolic fashion.
This because of (1). the density of states is large near the $M$
points, and only for the optical concentrations, the fermi surface
touches these regions, and (2) The momentum dependence of both the
inplane BCS pairing interaction(d-wave kind) and the interlayer pair
tunneling amplitude are such that they have large values near the $M$
also. So in a way the $T_c$ going through a maximum near the optimal
doping concentration, is expected for both the interlayer tunneling
mechanism and as well as for ordinary 
d-wave superconductor.
Now we shall discuss the variation of the buckling angle with the
doping concentration.

Some other feature of the WHA mechanism are,
 (1). The gap values near the $\Gamma-X$($0,\pm \pi$) points are
very much temperature
insentive at low temperatures, and drops rapidly to zero near the
$\rm{T}_c$. It is also
very sensitive to small doping variations, near the optimal doping
concentrations, i.e as the fermi surface moves away from the
$\Gamma-X$ points
because it cannot then take advantage of the large pair
tunneling amplitudes near those points in the BZ.
(2). The gaps along the diagonal direction in the BZ($\Gamma-M$ directions)
are on the other hand very sensitive to temperature variations,
falling
faster than BCS with temperature. It is also insensitive to small
doping variations near the optimal doping concentrations.
We shall discuss the possible consequences of these features of the
WHA gap later.

\section{Buckling of copper oxygen planes}

 The interlayer pair hopping term for a bilayer system is,
 $$
 \sum_{k}{t_{\perp}^2(k)\over t}
(c_{k\uparrow}^{1\dagger}c_{-k\downarrow}
 ^{1\dagger}c_{-k\downarrow}^2c_{k\uparrow}^2 + h.c. )
 $$
 Now the interlayer hopping matrix element $t_{\perp}$
 will be modulated due to the vibration out of plane of the
 oxygen atoms. We emphasize here again, that for the $B_{1g}$ mode
 the atoms all vibrate out of the CuO planes but the vibrations in two
 adjacent layers are in phase and hence the effective tunneling matrix
element
 of the Zhang-Rice orbitals (a linear combination of the oxygen
$p$ orbitals) from plane to plane dont change, but for
the $A_{1g}$ mode $t_{\perp}$ will change. We can do a Taylor
expansion of $t_{\perp}$ in small oxygen atom displacements.
$t_{\perp}(u)=t_{\perp}(0)+\alpha ({t_{\perp}\over c})u$
where $c$ is the z direction layer separation and $\alpha$ is a
constant of the
order of one.
Putting this back in the interlayer term we get the extra piece,
$$
 {u^2\over c^2}  \sum_{k}{t_{\perp}^2(k)\over t}
(c_{k\uparrow}^{1\dagger}
c_{-k\downarrow}
 ^{1\dagger}c_{-k\downarrow}^2c_{k\uparrow}^2 + h.c. ) +
{2\alpha u\over c}  \sum_{k}{t_{\perp}^2(k)\over t}
(c_{k\uparrow}^{1\dagger}
c_{-k\downarrow}
 ^{1\dagger}c_{-k\downarrow}^2c_{k\uparrow}^2 + h.c. )
$$
With these modifications the total effective Hamiltonian for the
phonon
modes corresponding to oxygen vibration out of plane, will be given
by:
\begin{equation}
H_{eff} =\sum_{k} {m\over 2}\omega _{0k}^2 u^2 -
 {u^2\over c^2}  \sum_{k}{t_{\perp}^2(k)\over t}\langle
(c_{k\uparrow}^{1\dagger}
c_{-k\downarrow}
 ^{1\dagger}\rangle \langle c_{-k\downarrow}^2c_{k\uparrow}^2\rangle +
h.c. ) -
{2\alpha u\over c}  \sum_{k}{t_{\perp}^2(k)\over t}\langle
(c_{k\uparrow}^{1\dagger}
c_{-k\downarrow}
 ^{1\dagger}\rangle \langle c_{-k\downarrow}^2c_{k\uparrow}^2\rangle +
h.c. )
\end{equation}
 where we have taken the anomolous averages. As we can see that the
second term renormalises the frequency of the buckling mode phonon,
and the term linear 
 $u$  shifts the harmonic oscillators
without
 affecting its energy(this will correspond to static
displacement of the oxygen atoms causing buckling
of the $CuO_2$ planes). 

 The renormalised frequency of the buckling mode 
phonon $\tilde \omega_{0k}$ and the static displacement of the oxygen
atoms out of the plane $u_0$ is given by,
\begin{equation}
\tilde \omega_{0k}^2 = \omega_{0k}^2 -  {2{\rm A}\over
m\omega_{0k}^2 c^2}~~;~~~~~{\rm and }~~~~u_o = {2\alpha c {\rm A}\over mc
\omega_{0k}^2 - {\rm A}}~~.
\end{equation}
where,
${\rm A} = \sum_{k}{t_{\perp}^2(k)\over t}\langle
(c_{k\uparrow}^{1\dagger}
c_{-k\downarrow}
 ^{1\dagger}\rangle \langle c_{-k\downarrow}^2c_{k\uparrow}^2\rangle +
h.c. )$
 
We take $\omega_k^0$ to be equal
to 340 cm$^{-1}$ or 10.2 THz for $k=0$.
In Fig: 1 and 2, we show the doping variation of both the superconducting
$T_c$ and the buckling angle $\theta = {\rm tan}^{-}(2u_0/a)$ where
$a$ is the latice spacing(Cu-O-Cu distance).
We see that both $T_c$ and angle $\theta$ scales the same way with
the doping concentration.
 As we have seen before,
 due to the momentum dependence 
of the pair tunneling amplitude
 at the optimal doping concentration, large number of pairs 
can delocalize by tunneling from plane to planes(thereby reducing the
kinetic energy). The buckling mode phonon helps in this process.
That is why the buckling mode phonon frequency softens and also
develops a static buckling distortion.


\section{Phonon mode softening}

Now let us take a look at the phonon frequency softening below $T_c$
for different phonon modes . We shall specifically look at the
$A_{1g}$ and $B_{1g}$ modes( corresponding to out of phase and inphase
vibrations of the oxygen atoms at adjacent bilayers). 

  Superconductivity induced change in frequency and linewidths
  have been used to
  estimate the superconducting gap function. This is accomplished by
  determining the phonon self energy changes that occur below the
$T_c$.
  The imaginary part of the phonon self energy is proportional to the
  phonon linewidth (inverse lifetime) which is measured as a function
  of temperature. If the phonon energy is less than twice the gap
energy ,
  it cannot decay by pair breaking and thus a decrease in linewidth
  below $T_c$ can result because of a decrease in the number of
available
  decay channels. If the phonon energy is greater than the gap,
  it has sufficient energy to dissociate an electron pair and the
  number of possible decay routes can increase below $T_c$, resulting
  in an increase in linewidth.

  The real part of the self energy is proportional to the phonon
  energy and the superconductivity induced changes are determined from
  a measurement of the
  phonon frequency as a function of temperature below $T_c$.
  Qualitatively one expects that, for phonon energies below or near
  the gap , the electron phonon interaction will push the frequencies
  to smaller values. Conversely for larger energy phonons
  ($\omega > 2\Delta$) are expected to undergo a
hardening. 

%
The phonon self energy,
assuming a polarisation bubble with no vertex correction is given by,
\begin{equation}
\sum_{\gamma}(q,i\nu _n) ={T\over N}
\sum_{k}\sum_{m}\vert\lambda_{\gamma}
(k,k+q)\vert^2
{\rm Tr} ( \tau_3G[k+q,i(\omega_n+\nu_n)]\tau_3
G(k,i\omega_n))
\end{equation}
where $\tau_3$ is a pauli matrix, $G(k,i\omega_n)$ is the fully
interacting green's function, $\omega_n$ ($\nu_n$) is the
fermion(boson)
Matsubara frequencies, and $\lambda_{\gamma}(k,k+q)$ is the electron
phonon matrix element for scattering an electron of
momentum $k$ to $k+q$ with momentum
transfer $q$ to or from a phonon with branch index $\gamma$.
The standard electron
phonon interaction being,
\begin{equation}
H_{e-ph}~~=~~\sum_{k,q,\gamma,\sigma} \lambda_{k}^{\gamma}(q)
c_{k-q,\sigma}^{\dagger}c_{k,\sigma}(a_{q,\gamma}^{\dagger}
+a_{-q,\gamma})
\end{equation}
In BCS theory
the superconducting Green's function is,
$$
G(k,i\omega_n)=-{i\omega_n+\epsilon_k\tau_3 +\Delta_k\tau_1 \over
\epsilon_k^2+\omega_n^2 +\Delta_k^2}
$$
Substituting this in the earlier equation and taking $q=0$ we get,
$$
\sum_{\gamma}(i\nu_n)=-{4\over N} \sum_{k} {\rm tanh }({E_k\over 2T})
{\Delta_k^2\vert \lambda_{\gamma k}\vert^2 \over
E_k[(2E_k)^2 +\nu_n^2]}
$$
where $E_k=(\epsilon_k^2+\Delta_k^2)^{1/2}$ and $\lambda_{\gamma k}=
\lambda_{\gamma}(k,k)$.
Raman shift is the difference between the superconducting and normal
state self
energies, where normal state self energy can be extraced from the
above equaition
by putting $\Delta_k=0$.
One still has to analytically continue ( $i\nu_n\rightarrow
\nu+i\delta$
and the real part describes the frequency shift and the imaginary part
describes the change in phonon linewidth upon entering the
superconducting state.

The simplest method of evaluating it,is to numerically evaluate
it on the lattice after solving the self consistent gap equation.
We have done so in this paper. We borrow the estimates of
the electron phonon coupling function from the estimates by
Normand et al\cite{nor}, i.e. $t(u)=t(0)[1-2.03{u\over a}]$ where $t$
is the inplane hopping matrix element, $u$ is the small oscillation
amplitude of the oxygen atoms out of the plane, and $a$ is
the lattice parameter. Also for the $A_{1g}$ mode symmetry
($u_i^x=-u_i^y$ ) and for the $B_{1g}$ mode ( $u_i^x=u_i^y$ ).
 The electron phonon coupling function $\lambda_{k,k}$ for $A_{1g}$
and $A_u$
 phonons will be $\lambda_{kk}=4.06{t(0)\over a}\sqrt{{\hbar \over
2m\omega
 }}\gamma(k)$, where $\gamma(k)=\cos k_x \pm \cos k_y$ for $A_{1g}$
and $B_{1g}$ phonon mode.
 The constants $m$ and $\omega$ are the mass of the oxygen atom and
the bare phonon
 frequencies($q=0$) in the normal state for the
 $A_{1g}$ and $B_{1g}$(340 cm$^{-1}$ and 307 cm$^{-1}$) modes.
This contribution to phonon frequency softening is due to the
screening
by the charge carriers in the planes.
We have already calculated  the phonon mode
softening due
 to interlayer pair hopping  process(additional screening due
to pair tunneling process).

Both the contributions that shifts the phonon frequencis, the usual
self energy
contribution and the screening effect due to the pair propagation
along the c-axis are shown as a function of temperature for the
$A_{1g}$ and $B_{1g}$ phonon modes
in Fig.3 and 4.

For the $B_{1g}$ phonon the second contribution will be missing and is
plotted in Fig. 4.
We see that the pair tunneling contribution is much larger than that
of screening by inplane
charge carriers. Seecndly from the form of electron phonon coupling
function $\lambda_{kk}$ as
well as the pair hopping contribution(the $k$ dependence of
$t_{\perp}(k)$), it is clear that, the contributions to the frequency
renormalisation of $A_{1g}$ and $B_{1g}$ phonons comes mostly from the
$M(0,\pi)$ points, and from the
points along the diagonal respectively. This in the light of the WHA
gap nature
would mean that
the change in frequency accross the $\rm{T}_c$ will be more abrupt for the
$A_{1g}$ phonon, while $B_{1g}$
frequency change will be gradual. At lower temperatures $A_{1g}$
frequency willi
very sensitive to
small doping variations near the optimal filling, decreasing as the
fermi surface moves away from

In conclusion, we find that, both scaling of $T_c$ and the 
buckling angle with doping variation and large frequency softening
of the buckling mode phonon compared to other optical modes of
comparable frequencies, can be understood with the WHA mechanism.

In $La_{2-x-y}Sr_xNd_yCuO_4$ compounds, it is noticed\cite{Buchner}
that with increase of $Nd$ concentration, the buckling(or the tilting
of $Cu)_6$ octahedra) increases while the $T_c$ falls, i.e for the
same doping concentration, the $T_c$ falls with buckling angle.
It was suggested by Bonesteel, Rice and Zhang\cite{bone} that the
charge carriers gets coupled with the buckling angle via spin orbit
coupling. It was shown by Bonesteel\cite{bb}
that spin orbit scattering 
can be a strong pair breaker provided tye material is in a
structurallydisordered phase,in which the buckling distrotion is
coherent on small length scale but is completely disorded on large
length scales( i.e. a spatial variation of buckling angle unlike what
we have considered). It is possible that this series of materials have
a lot of unrelaxed structural disorder.

\vskip 0.5cm

Fig.1  Transition temperature versus doping concentration
in the interlayer tunneling model.
\vskip 0.5cm

Fig.2  Transition temperature versus doping concentration
in the interlayer tunneling model.
\vskip 0.5cm

Fig.3  Frequency versus Temperature
of $A_{1g}$ phonon for three different chemical potentials($\mu =$
-457.0,-450.0,-440.0 meV, i.e doping concentrations of 0.19, 0.18,
0.16).
The corresponding $T_c$'s are 92.5, 88.0 and
85.0 degrees.
The upper three curves shows the screening of inplane carriers(the
bare polarisation bubble)
, the middle ones are the contributions due to interlayer tunneling,
and the bottom ones are the
total contributions.
\vskip 0.5cm

Fig.4  Frequency versus Temperature
of $B_{1g}$ phonon for three different chemical potentials($\mu =$
-457.0,-450.0,-440.0 meV,
 i.e doping concentrations of 0.19, 0.18,
0.16).

\newpage

\begin{figure}
\centerline{\psfig{figure=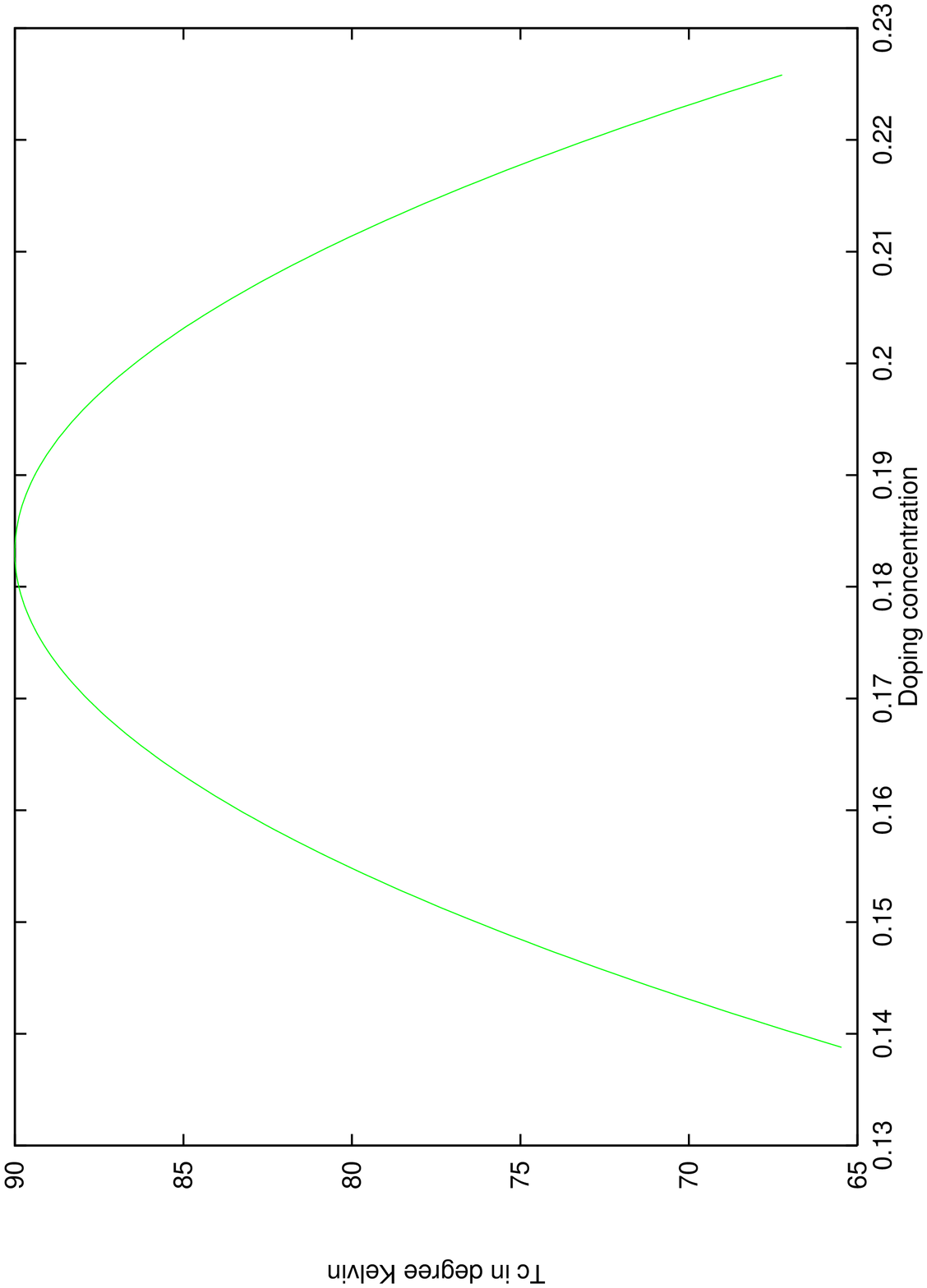}}
\end{figure}
\newpage

\begin{figure}
\centerline{\psfig{figure=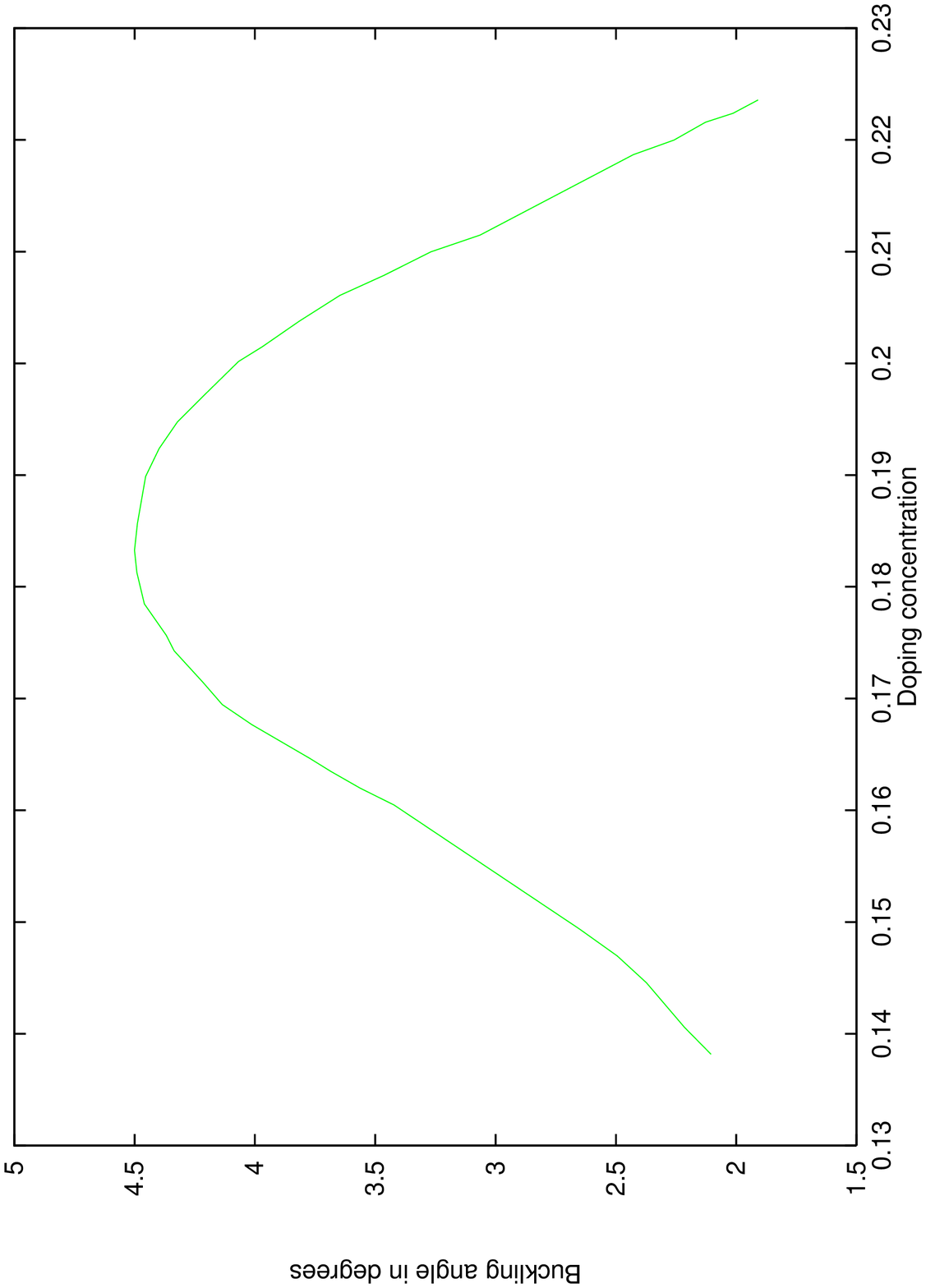}}
\end{figure}
\newpage

\begin{figure}
\centerline{\psfig{figure=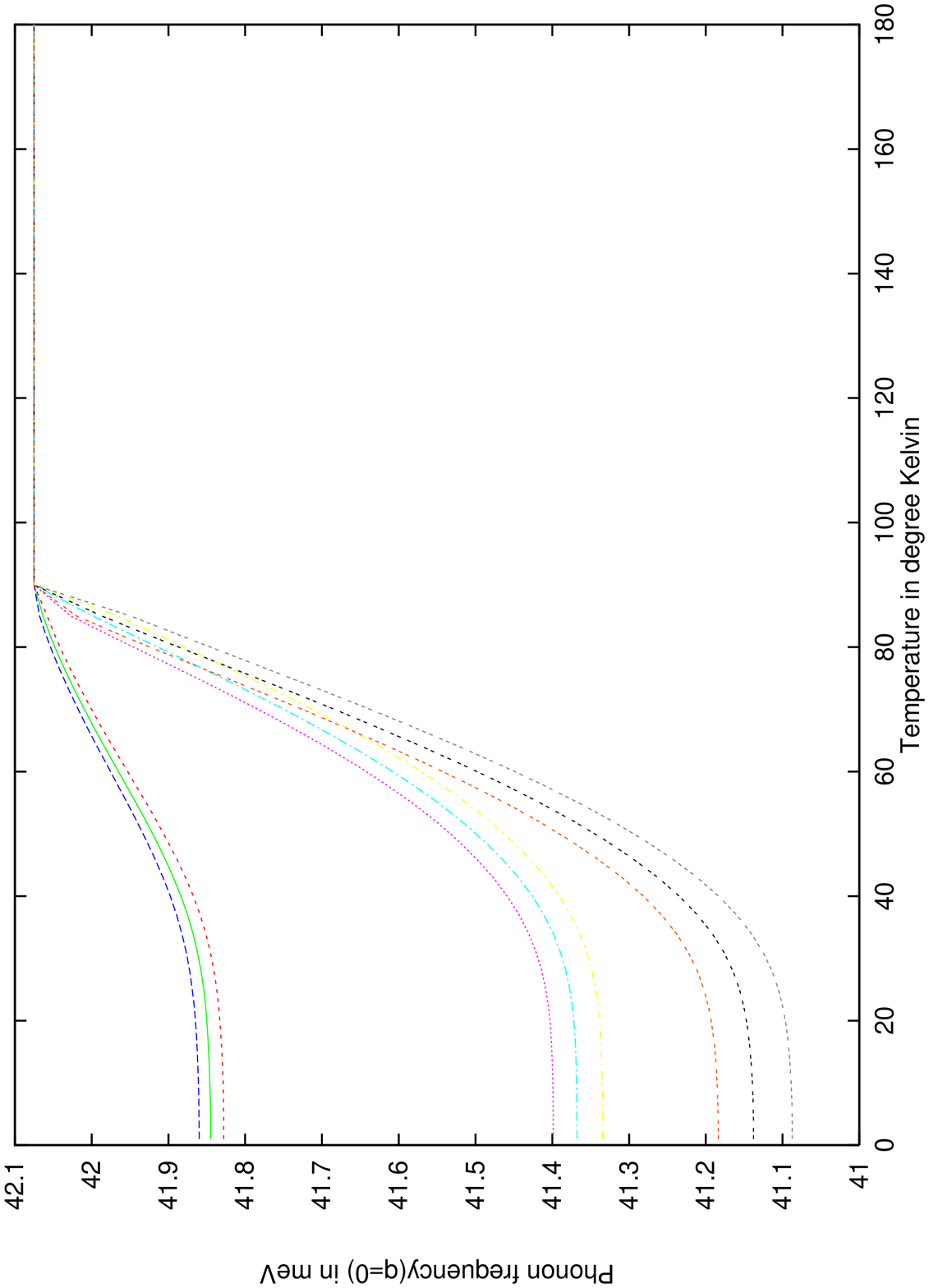}}
\end{figure}
\newpage

\begin{figure}
\centerline{\psfig{figure=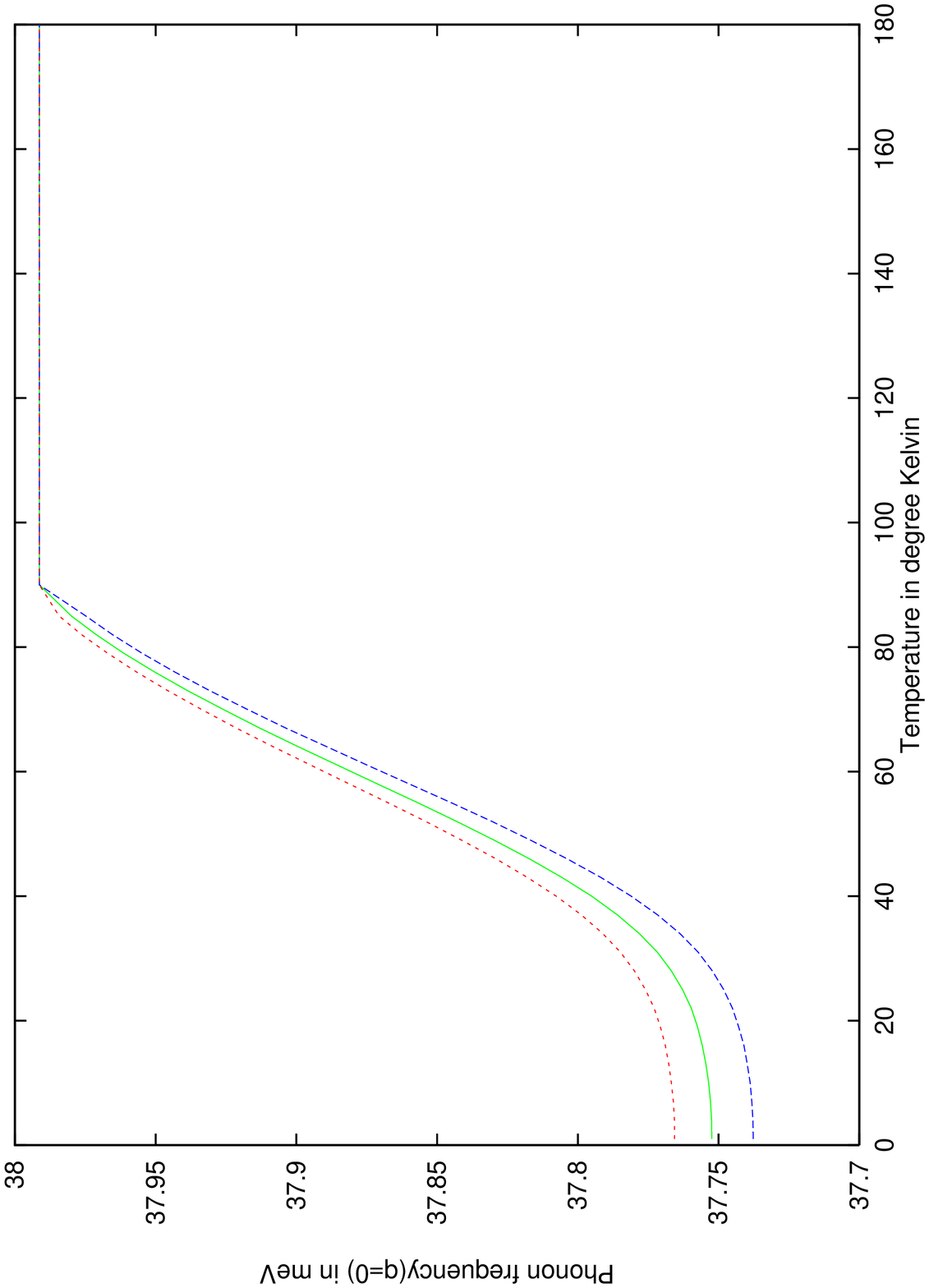}}
\end{figure}

\end{document}